\documentstyle[emulateapj]{article}

\lefthead{Nagao et al.}
\righthead{Mrk 573}

\begin{document}

\submitted{The Astronomical Journal, in press}
\title{Subaru Spectropolarimetry of Mrk 573:\\
       The Hidden High-Ionization Nuclear Emission-Line Region\\
       inside the Dusty Torus\footnote{
         Based on data collected at Subaru Telescope, which is operated
         by the National Astronomical Observatory of Japan.}
       }

\author{Tohru NAGAO$^{2,3}$, Koji S. KAWABATA$^4$, Takashi MURAYAMA$^3$, 
        Youichi OHYAMA$^5$, Yoshiaki TANIGUCHI$^3$,
        Yasuhiro SHIOYA$^3$, Ryoko SUMIYA$^3$, and Shunji S. SASAKI$^3$}

\vspace{5mm}

\affil{$^2$ INAF -- Osservatorio Astrofisico di Arcetri,
       Largo Enrico Fermi 5, 50125 Firenze, Italy;
       tohru@arcetri.astro.it}
\affil{$^3$ Astronomical Institute, Graduate School of Science, 
       Tohoku University, Aramaki, Aoba, Sendai 980-8578, Japan}
\affil{$^4$ Astrophysical Science Center, Hiroshima University, 
       1-3-1 Kagamiyama, Higashi-Hiroshima, Hiroshima 739-8526, Japan}
\affil{$^5$ Subaru Telescope, National Astronomical Observatory of
       Japan, 650 North A`ohoku Place, University Park, Hilo, HI 96720}


\begin{abstract}

We report on the result of our high quality spectropolarimetric 
observation for the narrow-line region in a nearby Seyfert 2 galaxy,
Mrk 573, by the Subaru Telescope.
The polarized flux spectrum of Mrk 573 shows not only prominent 
scattered broad H$\alpha$ emission but also various narrow 
forbidden emission lines.
We find that the measured polarization degree of the observed
forbidden emission lines is positively correlated with the 
ionization potential of the corresponding ions and the critical
density of the corresponding transitions.
We discuss some possible origins of these correlations, and then 
we point out that the correlations are caused due to the obscuration
of the stratified narrow-line region in Mrk 573 by the optically and
geometrically thick dusty torus, just similar to the previous 
study on NGC 4258.

\end{abstract}



\begin{deluxetable}{lcccc}
\scriptsize
\tablenum{1}
\tablecaption{Total Flux, Polarized Flux, and the Polarization Degree
              of Forbidden Emission Lines}
\tablewidth{0pt}
\tablehead{
\colhead{Line} &
\colhead{Total Flux\tablenotemark{a}} &
\colhead{Polarized Flux\tablenotemark{b}} &
\colhead{Pol. Degree} \\
\colhead{} &
\colhead{(10$^{-15}$ ergs s$^{-1}$ cm$^{-2}$)} &
\colhead{(10$^{-15}$ ergs s$^{-1}$ cm$^{-2}$)} & 
\colhead{(\%)} 
}
\startdata  
{}[O {\sc iii}]$\lambda$4959\tablenotemark{c} 
                           & 108.011$\pm$0.517 
                           & 1.161$\pm$0.007
                           & 1.1  \\
{}[O {\sc iii}]$\lambda$5007\tablenotemark{c} 
                           & 312.151$\pm$1.495
                           & 3.356$\pm$0.019
                           & 1.1  \\
{}[Fe {\sc vii}]$\lambda$6087
                           & 6.234$\pm$0.092 
                           & 0.116$\pm$0.007
                           & 1.9  \\
{}[O {\sc i}]$\lambda$6300\tablenotemark{c}   
                           &  7.490$\pm$0.113\tablenotemark{e}
                           & 0.102$\pm$0.006\tablenotemark{f}
                           & $<$1.4\\
\begin{tabular}{@{}l@{}}
{}[O {\sc i}]$\lambda$6364\tablenotemark{c} \\
{}[Fe {\sc x}]$\lambda$6374  
\end{tabular}
&
\begin{tabular}{@{}c@{}}
2.389$\pm$0.038 \\ 2.503$\pm$0.075
\end{tabular}
&
$\left\}\makebox{\rule[-2ex]{0ex}{2ex}{}}\right.$%
0.118$\pm$0.008\tablenotemark{g}
$\left\{\makebox{\rule[-2ex]{0ex}{2ex}{}}\right.$%
&
\begin{tabular}{@{}c@{}}
$<$1.4 \\ $>$3.4
\end{tabular} \\
{}[N {\sc ii}]$\lambda$6548\tablenotemark{c}  
                           & 23.439$\pm$0.365 
                           & 0.181$\pm$0.004
                           & 0.8  \\
{}[N {\sc ii}]$\lambda$6583\tablenotemark{c}  
                           & 69.145$\pm$1.077 
                           & 0.533$\pm$0.012
                           & 0.8  \\
\begin{tabular}{@{}c@{}}
{}[S {\sc ii}]$\lambda$6717\tablenotemark{d} \\ 
{}[S {\sc ii}]$\lambda$6731\tablenotemark{d}
\end{tabular}
&
\begin{tabular}{@{}c@{}}
17.514$\pm$0.364 \\ 20.844$\pm$0.408
\end{tabular}
&
$\left\}\makebox{\rule[-2ex]{0ex}{2ex}{}}\right.$%
0.316$\pm$0.014 
& 0.8\nl
{}[Ar {\sc iii}]$\lambda$7136
                           & 7.309$\pm$0.223
                           & 0.043$\pm$0.004
                           & 0.6  \\
\begin{tabular}{@{}c@{}}
{}[O {\sc ii}]$\lambda$7320\tablenotemark{d} \\ 
{}[O {\sc ii}]$\lambda$7330\tablenotemark{d}
\end{tabular}
&
\begin{tabular}{@{}c@{}}
1.609$\pm$0.062 \\ 2.012$\pm$0.069
\end{tabular}
&
$\left\}\makebox{\rule[-2ex]{0ex}{2ex}{}}\right.$%
0.049$\pm$0.005 
& 1.4
\enddata 
\tablenotetext{a}{Measured from the total flux spectrum corrected for
                  the contamination of the starlight of the host galaxy
                  by the manner described in the text.}
\tablenotetext{b}{Measured from the polarized flux spectrum on which
                  a 15 pixel smoothing is performed except for 
                  [N {\sc ii}]$\lambda \lambda$6548,6583, which are taken
                  from Nagao et al. (2004); see text.}
\tablenotetext{c}{The flux ratios and the velocity separations of the 
                  doublets of [O {\sc iii}], [O {\sc i}] and
                  [N {\sc ii}] are fixed to be the theoretical values.
                  Note that it is not the case for the [O {\sc i}]
                  doublet in the polarized flux spectrum (see text).}
\tablenotetext{d}{The velocity separations of the doublets of 
                  [S {\sc ii}] and [O {\sc ii}] are {\it not} fixed to 
                  be the theoretical values.}
\tablenotetext{e}{Excluding the contribution of the 
                  [S {\sc iii}]$\lambda$6312 emission.}
\tablenotetext{f}{Including the contribution of the polarized
                  [S {\sc iii}]$\lambda$6312 emission.}
\tablenotetext{g}{Sum of the polarized [O {\sc i}]$\lambda$6364 and
                  [Fe {\sc x}]$\lambda$6374 emission.}
\end{deluxetable}


\section{INTRODUCTION}

Active galactic nuclei (AGNs) can be broadly classified into two types;
type 1 AGNs and type 2 AGNs, based on the presence or absence of broad
permitted emission lines in their spectra. This dichotomy can be 
understood as follows: because high-velocity photoionized clouds in 
broad-line regions (BLRs) is surrounded by optically thick, 
dusty torus, the BLR can be seen only when we see the AGN from a 
face-on view toward dusty tori but is hidden by the torus when we see 
that from an edge-on view. This scheme is called AGN unified model
(see Antonucci 1993 for a review), 
which has been supported by various observational facts, e.g., 
the detection of hidden BLRs in some type 2 AGNs observed in the 
polarized light (e.g., Antonucci \& Miller 1985;
Miller \& Goodrich 1990; Tran, Miller, \& Kay 1992;
Young et al. 1993; Tran, Cohen, \& Goodrich 1995; Tran 1995; 
Young et al. 1996; Kay \& Moran 1998; 
Barth, Filippenko, \& Moran 1999a, 1999b; 
Tran, Cohen, \& Villar-Martin 2000; Kishimoto et al. 2001; Tran 2001;
Lumsden et al. 2001; Nagao et al. 2004).
This unified model is now one of the most important frameworks
of AGN studies.

Being different from the BLR emission, narrow forbidden and permitted 
emission lines are seen in spectra of both type 1 and type 2 
AGNs. This suggests that narrow-line regions (NLRs) are located far 
from the nucleus and thus its visibility should not depend on the 
viewing angle toward dusty tori. Recent imaging observations by HST 
indeed clarify the spatially extended feature of NLRs in the scale 
of $\sim$10$^{1-3}$ pc (e.g., Capetti et al. 1995; 
Schmitt \& Kinney 1996; Capetti, Axon, \& Macchetto 1997;
Falcke, Wilson, \& Simpson 1998; Ferruit et al. 1999; 
Schmitt et al. 2003). However, it has been recently recognized that 
high-ionization forbidden emission lines such as [Fe {\sc vii}] and 
[Fe {\sc x}], and high critical-density transitions such as 
[O {\sc iii}]$\lambda$4363 are statistically stronger in type 1 AGNs 
than in type 2 AGNs (e.g., Murayama \& Taniguchi 1998a; Schmitt 1998; 
Nagao, Taniguchi, \& Murayama 2000; Nagao, Murayama, \& Taniguchi 2001a,
2001b). Does this suggest the breakdown of the AGN unified model?
Murayama \& Taniguchi (1998b) showed that this observed 
differences can be understood in the framework of the unified model as 
follows: the highly-ionized dense gas clouds, which radiate
[O {\sc iii}]$\lambda$4363 and high-ionization lines selectively, 
are located near the nucleus where they can be hidden by the tori, 
as well as BLRs (see also Nagao et al. 2001a, 2001b).
Since this ``high-ionization nuclear emission-line region (HINER)'' 
plays a crucial role to construct realistic multi-zone photoionization
models for NLRs, it is very important to investigate the nature of 
this HINER component.

However, it is far from feasible to resolve such a small scale 
comparable to the inner radius of the tori ($<$ 1pc; e.g.,
Taniguchi \& Murayama 1998; see also Minezaki et al. 2004 and references
therein for the inferred inner radius of the tori by the dust
reverberation technique). This is the case where the 
spectropolarimetry can display its unique ability, as mentioned
already by Goodrich (1992).
Note that the spectropolarimetric study of NLRs has been
scarcely carried out up to now though many spectropolarimetric studies
have been performed for BLRs and the continuum emission. One of few 
exceptional AGNs whose NLR is extensively investigated in a 
spectropolarimetric manner is a LINER galaxy NGC 4258 (Barth et al. 1999c). 
This object exhibits the following three interesting aspects;
(1) the polarization degrees of forbidden lines are positively 
correlated with critical densities and ionization potentials of the 
line transitions, (2) the forbidden-line widths are broader in the
polarized spectrum than those in the total flux spectrum, and
(3) the gas density traced by the line ratio of the [S {\sc ii}] 
doublet is higher when using the polarized spectrum than when using 
the total flux spectrum. 
These observational facts are perfectly consistent with the HINER
hypothesis; that is, the above three results can be consistently 
explained by introducing relatively high velocity, high dense and highly 
ionized clouds, whose radiation contribute to the polarized spectrum 
significantly, in the innermost region of NLRs whose visibility 
depends on a viewing angle toward dusty tori. 
The same results are also obtained for a quasar-hosted ultraluminous 
infrared galaxy IRAS P09104+4109 (Tran et al. 2000). 
However unfortunately, NLRs in Seyfert galaxies, which are the most 
typical population of AGNs in the local Universe, have not yet been 
investigated in such a spectropolarimetric manner.

Therefore, we have started the project to perform spectropolarimetric
observations of NLRs in type 2 Seyfert galaxies by using the Subaru 
Telescope to explore the innermost of NLRs. Although many 
spectropolarimetric observations have been performed for type 2 Seyfert 
galaxies historically, their data quality is not high enough. 
This is because most of such previous studies aimed to search for 
scattered BLR emission, which can be detected more easily than the 
scattered components of weak forbidden lines. Since very high-quality 
spectropolarimetric data is required for our purpose, the large
aperture size of the Subaru Telescope is absolutely necessary.
In this paper, we present the initial result of our project on a 
Seyfert 2 galaxy, Mrk 573\footnote{The heliocentric radial velocity of 
Mrk 573 is 5156$\pm$90 km s$^{-1}$ (Whittle et al. 1988), which gives 
a projected linear scale of 0.33 $h_{75}^{-1}$ kpc for 1 arcsec.}.

\section{OBSERVATION AND DATA REDUCTION}

\begin{figure*}
\epsscale{1.50}
\plotone{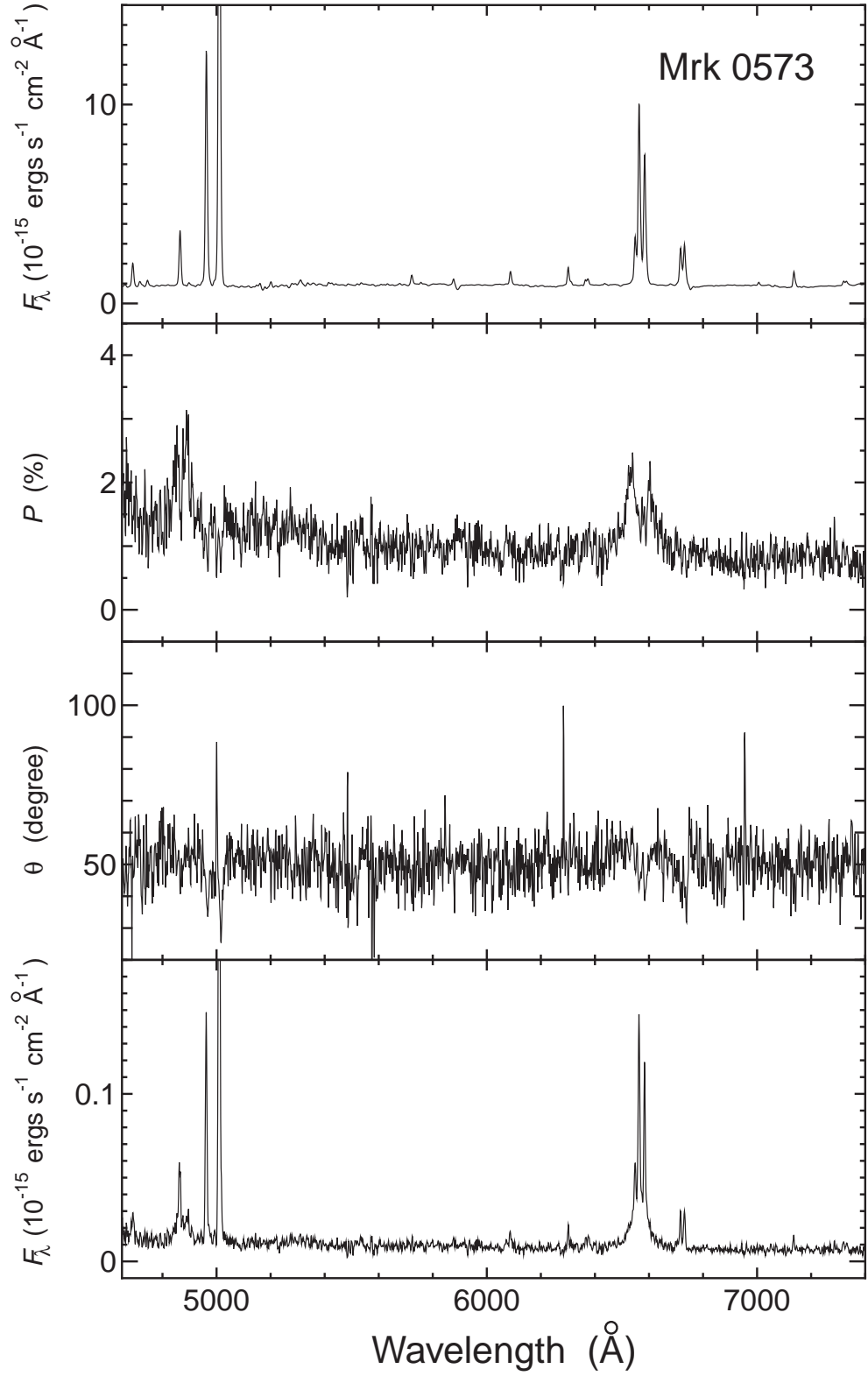}
\caption{
The obtained data of Mrk 573 are plotted as a function of wavelength.
The data are corrected for the interstellar polarization and the redshift
but not corrected for reddening and starlight of the host galaxy;
(a) the total flux, $I$,
(b) the polarization degree, $P$,
(c) the position angle of polarization, $\theta$, and
(d) the polarized flux (i.e., $I \times P$).
\label{fig1}}
\end{figure*}

\begin{figure*}
\epsscale{1.60}
\plotone{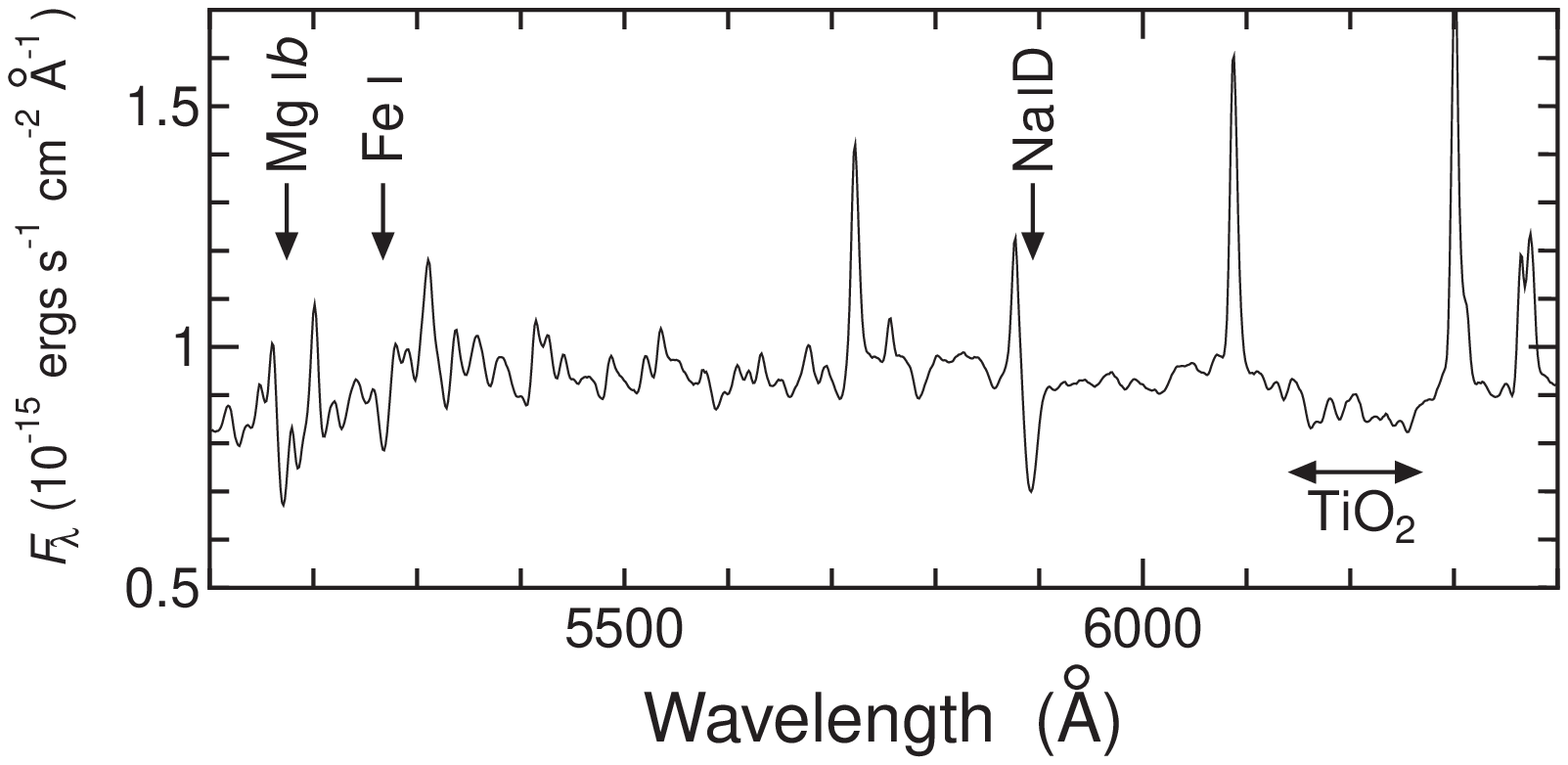}
\caption{
The enlarged total flux spectrum of Mrk 573.
Note that the starlight contribution from the host galaxy is not
corrected in this presented spectrum.
Some significant spectral features of the starlight are identified.
\label{fig2}}
\end{figure*}

The details of observations and the data reduction procedure are
basically described in Nagao et al. (2004). 
However, we briefly describe these issues below for the convenience
of the readers.

The spectropolarimetric observation for Mrk 573 was carried out by 
using FOCAS, Faint Object Camera And Spectrograph 
(Kashikawa et al. 2002) on the 8.2m Subaru Telescope (Kaifu et al. 2000;
Iye et al. 2004) at Mauna Kea, on 2003 October 5--6 (UT). 
All the observations were carried out through a polarimetric unit that 
consists of a rotating superachromatic half-wave plate and a quartz 
Wollaston prism. A 0$\farcs$4 width slit, 
a 300 lines mm$^{-1}$ grisms (300B), and an order-sorting filter of 
Y47 were used.
This setting results in a wavelength resolution of $R \sim 1000$.
We adopted a 3-pixel binning for the spatial direction on the chips,
which results in the spatial sampling rate of 0$\farcs$31 for a
binned pixel.
All of the data were obtained at four wave-plate position angles,
0.0$^{\circ}$, 45.0$^{\circ}$, 22.5$^{\circ}$, and 67.5$^{\circ}$.
The integration time of each exposure for the observation of Mrk 573
was 240 or 480 seconds, and the total on-source integration time was
208 minutes. The position angle of the slit was set to 0$^{\circ}$.
We also obtained spectra of unpolarized standard stars
(BD+28$^{\circ}$ 4211 and G191B2B) and a strongly polarized star (HD 204827).
Spectra of a halogen lamp and a thorium-argon lamp were also obtained
for the flat fielding and the wavelength calibration, respectively.

The data were reduced by the standard manner, by using IRAF\footnote{
IRAF (Image Reduction and Analysis Facility) is distributed by the
National Optical Astronomy Observatory, which is operated by the
Association of Universities for Research in Astronomy Inc., under 
corporative agreement with the National Science Foundation.}.
We extracted the spectra of Mrk 573 and the standard stars
by adopting the aperture size of 3$\farcs$1 (i.e., 10 binned pixels).
The corresponding linear aperture size in the frame of Mrk 573
is 1.02 $h_{75}^{-1}$ kpc $\times$ 0.13 $h_{75}^{-1}$ kpc.
The instrumental polarization was corrected by using the data of
the unpolarized standard stars. The instrumental depolarization was
not corrected because it has been confirmed experimentally that the
amount of the instrumental depolarization of the FOCAS is less than 
a few percent. The flux calibration was performed by using the data
of BD+28$^{\circ}$ 4211 and G191B2B (Oke 1990).
The polarization angle was calibrated by using the data of the strongly
polarized star.
The Galactic interstellar polarization toward the direction of
Mrk 573 is estimated to be $P=0.32$\% and $\theta=125\arcdeg$ at $B$ band,
based on the polarimetric properties of the two stars near the line of
sight toward Mrk 573; i.e., HD 9740 and HD 10441 (see Table 1 of
Nagao et al. 2004). 
Accordingly, the obtained spectrum was corrected for the Galactic 
interstellar polarization by adopting a
Serkowski law (Serkowski, Mathewson, \& Ford 1975) with $P_{\rm max}$ 
occurring at $\lambda_{\rm max}=5500{\rm \AA}$.

\section{RESULTS}

\begin{figure*}
\epsscale{1.40}
\plotone{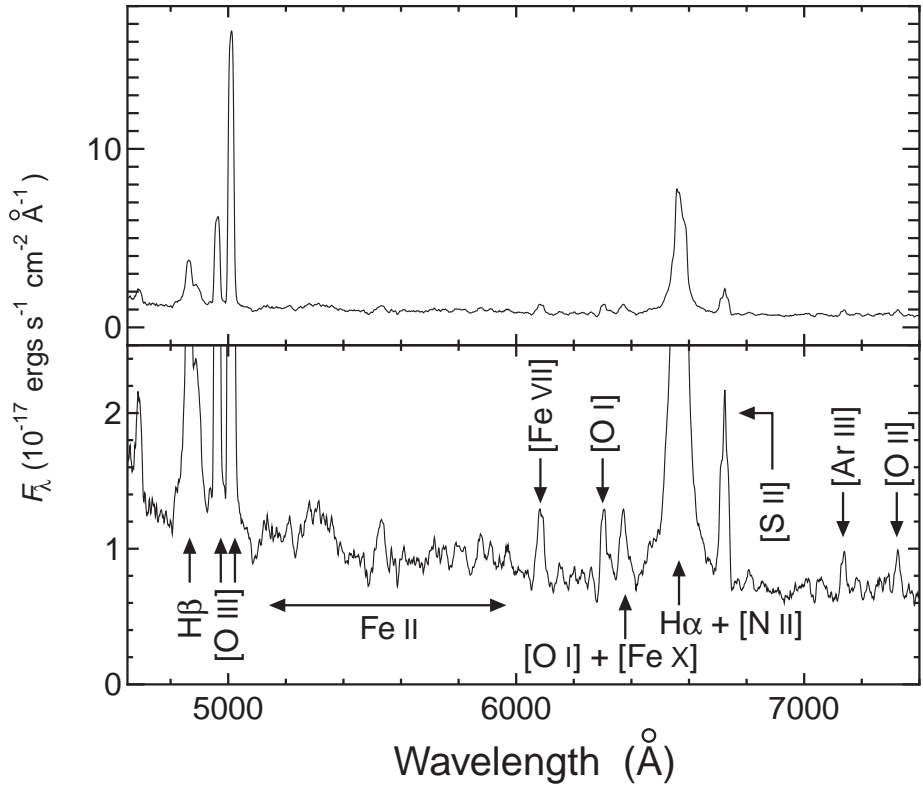}
\caption{
($Upper$) The spectrum of the polarized flux. A 15 pixel
smoothing is applied to improve the signal-to-ratio of the data.
($Lower$) Same as the upper panel, but the vertical scale is enlarged.
Emission-line features in the polarized flux spectrum are identified.
\label{fig3}}
\end{figure*}

The total flux ($I$), the polarization degree ($P$), the
position angle of polarization ($\theta$), and the polarized flux
($I \times P$) of Mrk 573 are shown as functions of wavelength
in Figure 1. The spectra shown in this figure are uncorrected for 
reddening and starlight of the host galaxy, but corrected for the
redshift and the Galactic interstellar polarization as described in \S2.
The total flux spectrum of Mrk 573 is significantly 
contaminated with the starlight of the host galaxy; e.g., Kay (1994)
reported that $\sim$80 \% of the continuum emission of Mrk 573 is
attributed to the starlight at 4400 ${\rm \AA}$.
In order to display how the obtained total flux spectrum of Mrk 573 is 
contaminated with the starlight, 
we show the enlarged spectrum of the total flux in Figure 2. 
The strong stellar features such as Mg {\sc i}$b$ $\lambda$5177,
Fe {\sc i} $\lambda$5270, Na {\sc i}D $\lambda$5893, and
TiO$_2$ molecular band at $\lambda$6235 are clearly seen.
Thus it should be kept in mind that the spectrum of the polarization 
degree shown in Figure 1b ($P = 1.0 \pm 0.2 \%$ at 
$5400 - 6000 {\rm \AA}$; see Nagao et al. 2004) is heavily diluted
by the unpolarized starlight emission.

As clearly exhibited in Figures 1b and 1d, a prominent broad
component of the H$\alpha$ emission is detected in the spectrum of
the polarized flux although there is no corresponding broad H$\alpha$
feature in the spectrum of the total flux (Figure 1a).
We do not discuss this hidden-BLR emission in this paper, because 
this issue is discussed in the companion paper (Nagao et al. 2004)
in detail. In addition to the broad emission lines, 
strong forbidden lines such as
[O {\sc iii}]$\lambda \lambda$4959,5007,
[N {\sc ii}]$\lambda \lambda$6548,6583, and
[S {\sc ii}]$\lambda \lambda$6717,6731 are clearly seen in the
polarized flux spectrum and thus the polarization degree of
such strong forbidden emission lines can be
measured rather easily. However, being different from such strong
forbidden emission lines, there are two difficulties to measure
polarization degrees for rather weak but important forbidden lines.

One is the insufficient signal-to-noise ratio of the polarized flux
spectrum. It is generally very difficult to obtain high quality 
spectra of polarized flux even by using 8m class telescopes, because
it requires a huge number of photons.
In order to improve this situation, we apply a 15 pixel
(which corresponds to $\approx$20.9 ${\rm \AA}$) smoothing on the
spectrum of the polarized flux. The resulting spectrum and its enlarged
one are shown in Figure 3. This smoothed spectrum clearly shows
various emission-line features including the Fe {\sc ii} multiplet,
[Fe {\sc vii}]$\lambda$6087, [Fe {\sc x}]$\lambda$6374,
[Ar {\sc iii}]$\lambda$7136, and 
[O {\sc ii}]$\lambda \lambda$7320,7330.
The [Fe {\sc x}]$\lambda$6374 emission line is blended with the
[O {\sc i}]$\lambda$6364 emission in the smoothed spectrum.
However, we can recognize that the [Fe {\sc x}]$\lambda$6374 emission
is significantly polarized, because we can estimate the contribution
of [O {\sc i}]$\lambda$6364 in the blended line by assuming the
theoretical flux ratio of
[O {\sc i}]$\lambda$6300/[O {\sc i}]$\lambda$6364 $\approx$ 3.14.
We can see that the strength of the blend of
[O {\sc i}]$\lambda$6364 and [Fe {\sc x}]$\lambda$6374 is
significantly stronger than one third of that of [O {\sc i}]$\lambda$6300
in the polarized flux spectrum as shown in Figure 3.
The [O {\sc ii}]$\lambda \lambda$7320,7330 doublet is also blended in
the smoothed spectrum of the polarized flux.

The other difficulty for the measurement of polarization degrees of
weak emission lines is the contamination of the starlight from the
host galaxy into the total flux spectrum. Since the continuum emission
of the total flux of Mrk 573 is heavily contributed from the
host galaxy as mentioned above (see also Kay 1994), the fluxes of
weak emission lines are hard to be measured accurately owing to
the complex spectral features of the starlight.
Thus we should remove the starlight contribution in the total flux
spectrum by using the model spectrum, in order to measure the
fluxes of the weak emission lines in the total flux spectrum correctly.
The stellar population of a host galaxy of AGNs are generally hard 
to be determined accurately (see, e.g., 
Schmitt, Storchi-Bergmann, \& Cid Fernandes 1999; Raimann et al. 2003). 
However, we are not interested in the accurate determination
of the stellar population of the host galaxy of Mrk 573 but only
interested in removing the stellar features roughly, in order to
measure the polarization degrees of emission lines.
We thus try to remove the stellar features by
using only some template galaxy spectra.

\begin{figure*}
\epsscale{1.10}
\plotone{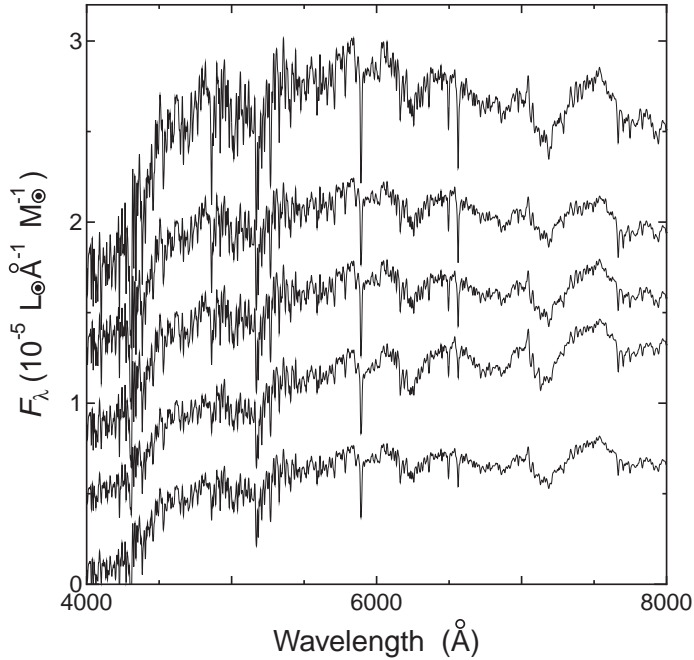}
\caption{
Spectra of five model SSPs of Bruzual \& Charlot (2003)
examined as a template of the host galaxy of Mrk 573 in our analysis.
The spectra are smoothed to match the spectral resolution of
our observational data, $R = 1000$.
The adopted parameters for the SSPs are (Age, $Z$) =
(5 Gyr, 1.0 $Z_{\odot}$), (10 Gyr, 0.4 $Z_{\odot}$),
(10 Gyr, 1.0 $Z_{\odot}$), (10 Gyr, 2.5 $Z_{\odot}$), and
(15 Gyr, 1.0 $Z_{\odot}$), which are presented from top to bottom
in the figure, respectively.
The SSPs are displayed in unit of the solar luminosity per angstrom
except for the model with (Age, $Z$) = (15 Gyr, 1.0 $Z_{\odot}$),
for which the constant of $5 \times 10^{-6} L_{\odot} {\rm \AA}^{-1}$
is subtracted for the presentation.
\label{fig4}}
\end{figure*}

As for the model spectrum, we adopt the simple stellar population (SSP)
of Bruzual \& Charlot (2003). 
We examine five SSPs as candidates of the representative stellar
population of the host galaxy of Mrk 573; that is, (Age, $Z$) =
(5 Gyr, 1.0 $Z_{\odot}$), (10 Gyr, 0.4 $Z_{\odot}$),
(10 Gyr, 1.0 $Z_{\odot}$), (10 Gyr, 2.5 $Z_{\odot}$), and
(15 Gyr, 1.0 $Z_{\odot}$).
Since the spectral resolution of the SSP spectra ($\approx$2000; see
Bruzual \& Charlot 2003) is not the same as
that of the data obtained by our observation, we use the Gaussian kernel
to match the spectral resolution of the models to the FOCAS data,
$R = 1000$. The smoothed spectra of the five SSPs are shown in Figure 4.
In order to judge which SSP is better as a template of the starlight
of the host galaxy, we focus on two stellar absorption features;
Na {\sc i} D $\lambda$5893 and the TiO$_2$ molecular band at
$\approx$6235 ${\rm \AA}$ (in the rest frame).
Among the five SSPs, those with (Age, $Z$) = (5 Gyr, 1.0 $Z_{\odot}$),
(10 Gyr, 0.4 $Z_{\odot}$), and (10 Gyr, 1.0 $Z_{\odot}$) are not
appropriate for the host galaxy spectrum. This is because the
equivalent widths of Na {\sc i} D $\lambda$5893 and the TiO$_2$
molecular band of these three SSPs are too small. In the remaining two SSPs,
the one with (Age, $Z$) = (10 Gyr, 2.5 $Z_{\odot}$) cannot eliminate the
absorption features of Na {\sc i} D $\lambda$5893 and the TiO$_2$
molecular band, simultaneously.
We thus finally adopt the SSP with (Age, $Z$) =
(15 Gyr, 1.0 $Z_{\odot}$) as the template spectrum of the host galaxy
of Mrk 573. The close-up views of the total flux spectra
before and after the subtraction of the model galaxy spectrum are shown
in Figure 5. The whole of the total flux spectrum after
subtracting the starlight is shown in Figure 6.
The fraction of the starlight of the host galaxy in the observed
continuum spectrum of the total flux is $\approx$84\% at
$\sim$5500${\rm \AA}$, which is roughly consistent with the result of
Kay (1994).

\begin{figure*}
\epsscale{1.60}
\plotone{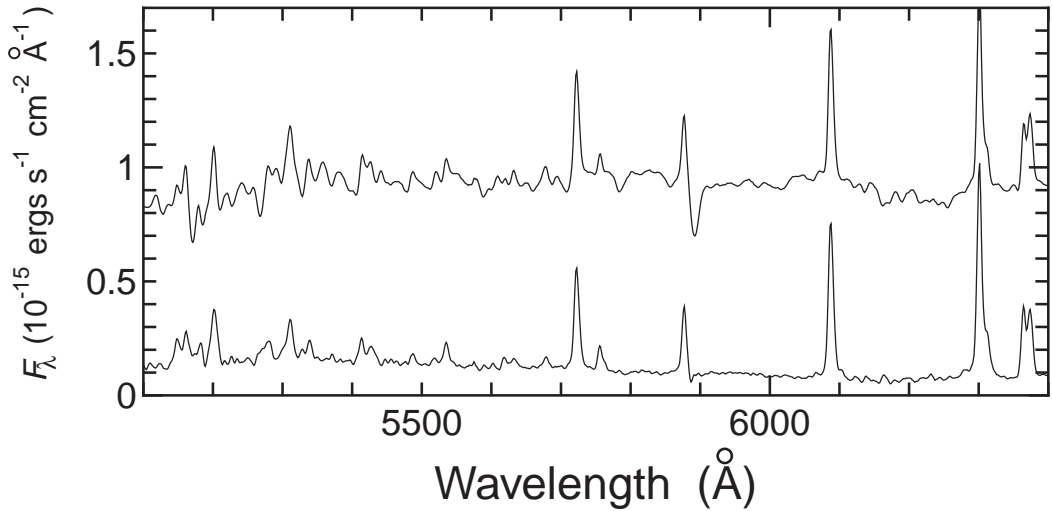}
\caption{
Close-up views of the total flux spectra
before and after the subtraction of the model galaxy spectrum,
i.e., the SSP of Bruzual \& Charlot (2003) with (Age, $Z$) = 
(15 Gyr, 1.0 $Z_{\odot}$).
\label{fig5}}
\end{figure*}

After these two treatments of the data to improve the measurement
accuracy, we are now ready to measure the polarization degree of weak
emission lines of Mrk 573.
In the process of the measurements for the emission-line fluxes,
we use the task SPECFIT in the IRAF, which was developed by
Kriss (1994), assuming the Gaussian profile for the emission lines.
We measure the unpolarized fluxes of the emission lines by using
the total flux spectrum after removing the contribution of the starlight
of the host galaxy. The flux ratios and the velocity separations
of the doublets of [O {\sc iii}], [O {\sc i}] and [N {\sc ii}] are
fixed to be the theoretical values\footnote{Here we adopt the flux ratios 
of 2.89, 0.319, and 2.95, and the wavelength ratios of 
1.00967, 1.01007, and 1.00539, for the doublets of 
[O {\sc iii}]$\lambda\lambda$4959,5007,
[O {\sc i}]$\lambda\lambda$6300,6364, and 
[N {\sc ii}]$\lambda\lambda$6548,6583, respectively
(see, e.g., Osterbrock 1989).}.
As for the doublets of [S {\sc ii}] and [O {\sc ii}], however,
the velocity separations are not fixed to be the theoretical values,
because the two emission lines in each doublet generally arise
at different clouds in the NLR (see, e.g., Ferguson et al. 1997).
For the measurement at the wavelength region around
[O {\sc i}] $+$ [Fe {\sc x}], we include the [S {\sc iii}]$\lambda$6312
emission in the multi-Gaussian spectral modeling process because
the [O {\sc i}]$\lambda$6300 emission is blended by that emission
as seen in Figure 3. The measured flux ratio of 
[S {\sc iii}]$\lambda$6312/[O {\sc i}]$\lambda$6300 is $\sim$0.26.
As for the polarized emission-line spectrum, the measurement is
performed by using the polarized flux spectrum smoothed by 15 pixels
except for the [N {\sc ii}] doublet, which cannot be measured in
the smoothed spectrum due to the blending with the H$\alpha$ emission.
For the [N {\sc ii}] flux, we thus refer the value presented in
Nagao et al. (2004), in which no smoothing was performed for
the polarized flux spectrum.
The adjacent pairs of emission lines,
[O {\sc i}]$\lambda$6300 plus [S {\sc iii}]$\lambda$6312,
[O {\sc i}]$\lambda$6363 plus [Fe {\sc x}]$\lambda$6374,
[S {\sc ii}]$\lambda$6717 plus [S {\sc ii}]$\lambda$6731, and
[O {\sc ii}]$\lambda$7320 plus [O {\sc ii}]$\lambda$7330
cannot be deblended in the smoothed
spectrum. Thus we adopt a single Gaussian profile for each pair,
for the rough measurement of the sum of the emission-line fluxes.

The measured total flux and polarized flux, and the polarization
degree for each forbidden emission line are given in Table 1.
The derived polarization degree of the [O {\sc iii}] emission 
is higher than the previously reported value, 0.27$\pm$0.09 \%
(Goodrich 1992). This may be due to the difference in the 
aperture size to make the one-dimensional spectra.
Note that the slit width that we adopted is far narrower than
that of the observation of Goodrich (1992), 2$\farcs$0 arcsec.
The upper-limited value of 1.4\% for the polarization degree of
[O {\sc i}]$\lambda$6300 is obtained by assuming that 
the [S {\sc iii}]$\lambda$6312 emission is unpolarized.
Under this assumption, we can estimate the polarized flux of the
[Fe {\sc x}]$\lambda$6374 emission to be $8.5 \times 10^{-17}$
ergs s$^{-1}$ cm$^{-2}$, which corresponds to the polarization
degree of 3.4\%.
On the other case, if the flux ratio of 
[S {\sc iii}]$\lambda$6312/[O {\sc i}]$\lambda$6300
in the polarized flux spectrum is the same as that in the
total flux spectrum, the polarized flux of [O {\sc i}]$\lambda$6300
would be $8.1 \times 10^{-17}$ ergs s$^{-1}$ cm$^{-2}$.
In this case, the polarization degrees of [O {\sc i}] and [Fe {\sc x}]
are estimated to be 1.1\% and 3.7\%, respectively.

\section{DISCUSSION}

\begin{figure*}
\epsscale{1.60}
\plotone{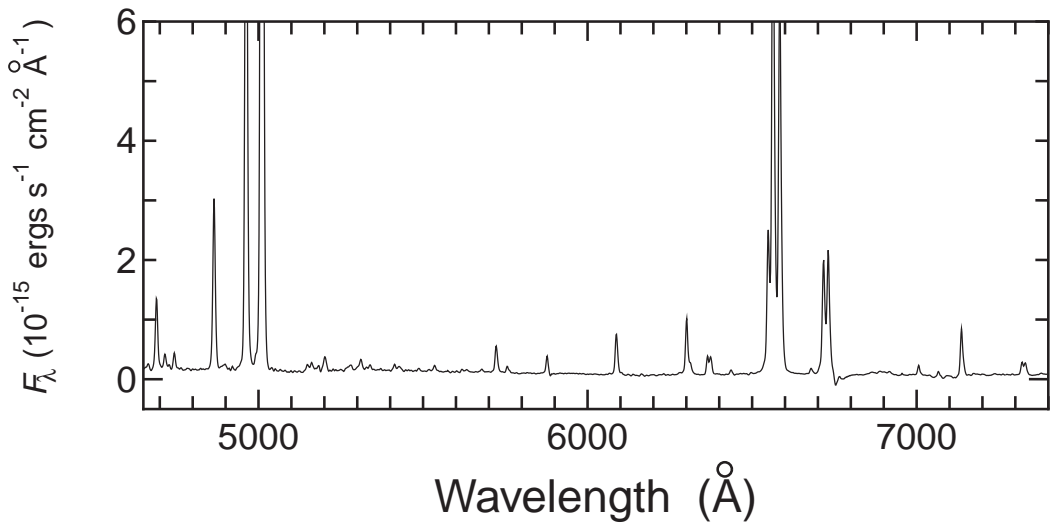}
\caption{
Total flux spectrum of Mrk 573 after subtracting the starlight
contribution from the host galaxy, shown in the whole range of the
spectral coverage.
\label{fig6}}
\end{figure*}

In order to interpret the obtained results, we plot the polarization
degrees of the forbidden emission lines as functions of
the ionization potential of corresponding ions and the
critical density of corresponding transitions,
in Figure 7. As shown in Figure 7, positive correlations are seen 
both between the polarization degree and the ionization potential 
and between the polarization degree and the critical density.
Note that the typical error for the polarization degree is 0.1\% 
at most, for faint emission lines.
These correlations appear to be very similar to those seen in
NGC 4258 (Barth et al. 1999c) and IRAS P09104+4109 (Tran et al. 2000).
To investigate whether or not these positive correlations are
statistically significant, we apply Spearman's rank test
where the null hypothesis is that the obtained polarization degree
is not correlated with the ionization potential and with the critical
density. The calculated probabilities that the data is consistent
with the null hypothesis are 0.313 and 0.038 for the correlations
of the polarization degree on the ionization potential and
the critical densities respectively, in the case that
the [S {\sc iii}]$\lambda$6312 emission is not polarized.
As for the case that the polarization degree of 
the [S {\sc iii}]$\lambda$6312 emission is the same as the
[O {\sc i}]$\lambda$6300 emission, the probabilities are calculated
to be 0.213 and 0.033, respectively.
These results suggest that the correlation between the polarization
degree and the ionization potential is statistically insignificant
although that between the polarization degree and the critical
density is statistically significant.
This tendency seems to resemble the situation seen in NGC 4258 
(see Figure 3 of Barth et al. 1999c).
This result does not depend strongly on the assumption about
the polarization property of the [S {\sc iii}]$\lambda$6312 emission.
Note that if the [S {\sc iii}]$\lambda$6312 emission is more polarized 
than the [O {\sc i}]$\lambda$6300 emission, the [O {\sc i}]$\lambda$6300 
and [Fe {\sc x}]$\lambda$6374 polarization are correspondingly estimated 
to be $< 1.1$\% and $> 3.7$\%, respectively. 
Even in this case, the correlations seen in Figure 7 do not disappear. 
However, the [S {\sc iii}]$\lambda$6312 emission may be not so strongly 
polarized because its critical density ($\sim1 \times 10^6$ cm$^{-3}$) 
is lower than the critical density of the [O {\sc i}]$\lambda$6300 
emission ($\sim2 \times 10^6$ cm$^{-3}$).

We then discuss the origin of the correlations.
Sometimes NLR emission lines look to be polarized due to
the Galactic interstellar polarization; i.e., narrow emission lines
without intrinsic polarization can be seen in the polarized flux 
spectrum when the Galactic interstellar polarization is not
corrected properly (e.g., Young et al. 1996).
If the NLR polarization and its dependences on the ionization potential
and the critical density are significantly attributed to the improper 
correction of the Galactic interstellar polarization, the polarization 
degree of the lines should be a function of the wavelength. 
Thus we plot the polarization degrees of the forbidden emission
lines as a function of the line wavelength, in Figure 8.
As shown in this figure, the polarization degree is not correlated
with the line wavelength. This result strongly suggests that the 
observed NLR polarization and the correlations seen in Figure 7 are 
intrinsic, not due to the Galactic interstellar polarization.
No clear correlation between the line flux and the
polarization degree (Table 1) is also consistent with this idea.

The correlations may be created by the nuclear star-forming activity,
because H {\sc ii} regions radiate lower ionization emission lines
selectively than NLRs in AGNs. Since the emission from nuclear
star-forming regions is thought to show a very small or no polarization,
the observed polarization degree of lower-ionization emission lines 
may be more diluted by the nebular emission of the star-forming regions,
than that of higher-ionization emission lines.
And accordingly, the dependence of the dilution on the ionization 
potential of the forbidden lines could cause the observed correlations 
shown in Figure 7. However, this is not the probable scenario for Mrk 573.
This is because the spectral properties of the host galaxy of Mrk 573
suggest that the stellar population of the host galaxy is very old
and do not contain significant amount of young stars,
which is suggested by the result of our SSP fitting (\S3).
Gonz\'{a}lez Delgado, Heckman, \& Leitherer (2001) also reported
that the stellar emission of Mrk 573 is dominated by a
old stellar population component, by a more quantitative manner
than ours (see also Cid Fernandes et al. 2001; Raimann et al. 2003).
Mouri \& Taniguchi (2002) reported that the far-infrared emission
of Mrk 573 is dominated by the AGN emission, not by the nuclear or 
circumnuclear starburst activity, which also supports the idea that 
the NLR emission of Mrk 573 is not diluted so significantly by the
emission attributed by the star-forming activities.

\begin{figure*}
\epsscale{0.90}
\plotone{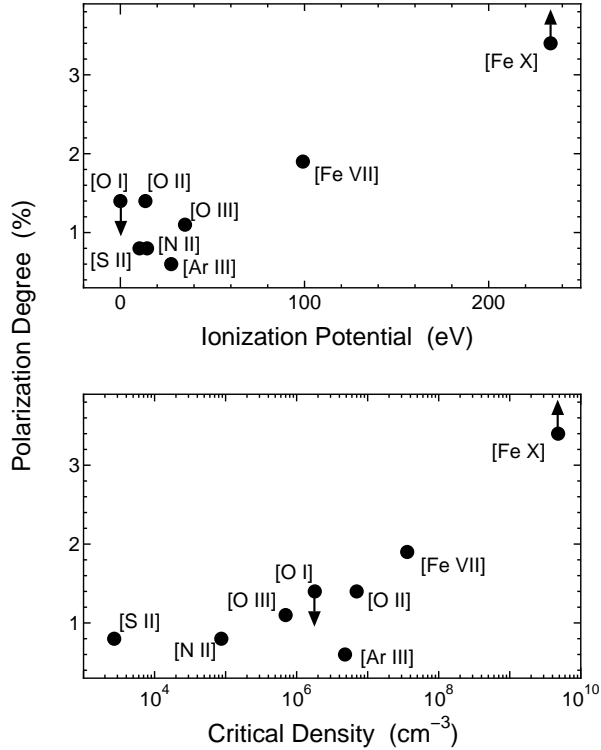}
\caption{
Polarization degrees of the forbidden emission lines as
functions of the ionization potential of corresponding ions 
and the critical density of corresponding transitions.
The polarization degrees of [O {\sc i}] and [Fe {\sc x}] are
upper and lower limited values, respectively (see text).
\label{fig7}}
\end{figure*}

A possible origin of the correlations inferred from the AGN unified
model is that the highly ionized, dense part of NLRs (i.e., HINERs) is 
hidden by an edge-on, geometrically and optically thick torus.
If the NLRs possess a stratified structure, highly ionized and/or
high dense gas clouds should be more effectively hidden by the thick torus.
In this case, emission lines with a higher ionization potential and/or
a higher critical density can be seen in the polarized flux spectrum
much more, which results in the positive correlations of
the polarization degree of the forbidden emission lines with
the ionization potential and the critical density.
Since this scenario has been discussed for IRAS 20210+1121
(Young et al. 1996), NGC 4258 (Barth et al. 1999c), and 
IRAS P09104+4109 (Tran et al. 2000), it is very interesting to
investigate whether or not this scenario is also the case for Mrk 573.

First, we focus on how significant the HINERs are hidden by the dusty 
torus in Mrk 573. Nagao et al. (2001a) demonstrated that the 
emission-line diagnostic diagram of 
[O {\sc iii}]$\lambda$4363/[O {\sc iii}]$\lambda$5007 vs.
[Fe {\sc vii}]$\lambda$6087/[O {\sc iii}]$\lambda$5007 is useful to
examine the obscuration of HINERs (see Figure 12 of Nagao et al. 2001a).
The optical spectrum of Mrk 573 shows only weak HINER emission
([O {\sc iii}]$\lambda$4363/[O {\sc iii}]$\lambda$5007 $\sim$ 0.018 and
[Fe {\sc vii}]$\lambda$6087/[O {\sc iii}]$\lambda$5007 $\sim$ 0.013;
Nagao et al. 2000, 2001a), which is diagnosed that the HINER in
Mrk 573 is very obscured compared to other Seyfert galaxies.
Second, we refer to the previous literature about the spatially resolved
observations of the NLRs. 
As for the HINER emission, Winge et al. (2000) presented by their
near-infrared spectroscopy that the [S {\sc ix}]1.252$\mu$m emission is 
unresolved although the neighboring [Fe {\sc ii}]1.257$\mu$m emission is
clearly extended in Mrk 573. Low-ionization emission lines such
as [O {\sc iii}]$\lambda$5007 are, on the contrary, clearly extended 
around the central engine of Mrk 573 (e.g., Unger et al. 1987; 
Haniff, Wilson, \& Ward 1988; Tsvetanov \& Walsh 1992;
Pogge \& De Robertis 1995; Capetti et al. 1996; Schmitt et al. 2003).
Since the polarization of the emission from this extended NLR is thought
to be very small, the polarization degrees of low-ionization emission
lines are diluted and thus the observed correlations can be created.
Third, we check where the polarization of the NLR emission occurs
in Mrk 573. It is known that the polarization angle of the nuclear 
continuum emission of type 2 AGNs tends to be perpendicular to the 
direction of extended NLRs and radio jets (e.g., Antonucci 1983; 
Brindle et al. 1990). This is attributed by the fact that the 
polarization is due to the scattering by free electrons in the opening 
cone above the torus. As for Mrk 573, the polarization angle of the
nuclear continuum emission obeys the same tendency as the other
type 2 AGNs (e.g., Nagao et al. 2004), which suggests that the observed 
polarization of the forbidden emission lines is also caused in the 
opening cone above the torus. Thus the scattering material can see
the obscured HINER emission, which enables the HINER emission
to be observed in the polarized flux spectrum.

If the polarization mechanism of the forbidden lines is the same 
as that of the broad permitted lines and the continuum emission, 
the position angle of the polarization of the forbidden lines 
should be similar to those of the broad lines and continuum 
emission. But, as expected from Figure 1c, the S/N of our data is 
not so sufficient to compare independently the position angles of 
the weaker forbidden lines, although some strong forbidden lines 
may show somewhat different position angles (Figure 1c).
It should be thus examined by another deeper observation.

Taking all of the above matters into account, we conclude that
the observed correlations presented in Figure 7 is caused due to
the obscuration of the stratified NLR of Mrk 573 by the thick dusty
torus, just similar to the previous studies on NGC 4258 and
IRAS P09104+4109. This result is important as an evidence of
the stratification of NLRs in AGNs. The velocity profiles of 
polarized forbidden emission lines are more useful to investigate
this issue further (Barth et al. 1999c), but the quality and the velocity 
resolution of our data are not high enough to carry out such an analysis.

Here we discuss the possibility that the correlations seen in Figure 7
are caused without any obscuration effect. If NLRs are stratified and
thus HINERs reside in the innermost part of NLRs, photons arising at
HINERs move longer optical path in the scatterer than the photons
arising at the outer, low-ionization NLRs. The correlation of the 
polarization degree with the ionization potential and the critical
density may be attributed to this difference. If this effect is the
dominant source of the correlations seen in Figure 7 and the obscuration
by the torus is not so important to create the correlations,
the correlations should be discovered also in type-1 AGNs, just similar
to type-2 AGNs such as Mrk 573 and NGC 4258.
Further spectropolarimetry of NLRs in AGNs by 8m-class telescopes
will bring us useful information about the structure of the 
innermost part of NLRs, which is never resolved spatially by 
current observational facilities.

\begin{figure*}
\epsscale{1.50}
\plotone{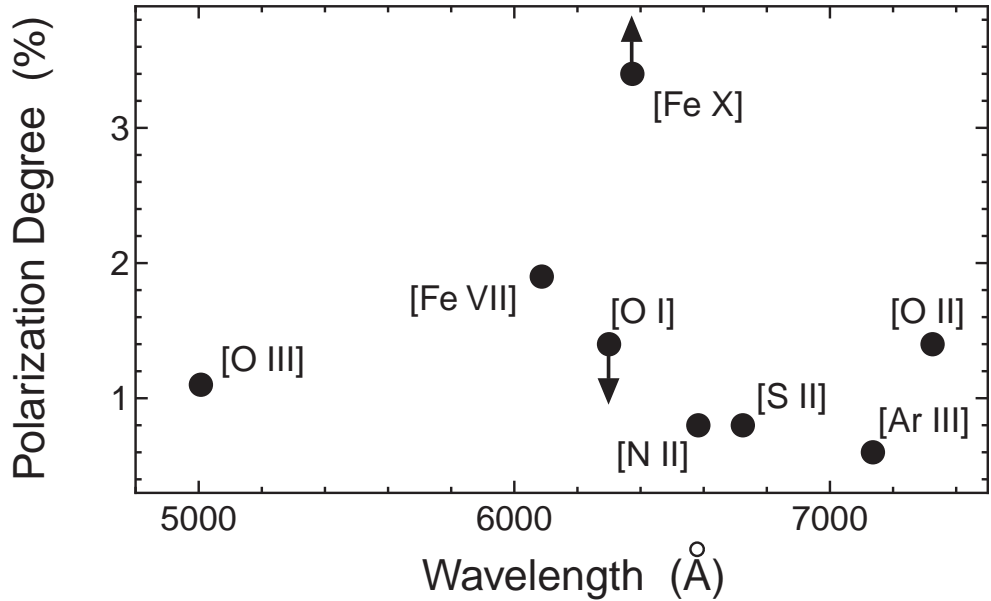}
\caption{
Polarization degrees of the forbidden emission lines as a function
of the rest wavelength of the emission lines. The arrows are the same
as those in Figure 7.
\label{fig8}}
\end{figure*}

Finally we remark that the measured polarization degrees of
the forbidden emission lines, $\sim$0.6\% -- 3.4\%, are comparable to or
larger than the polarization degree of the continuum emission
($\sim$1.0 \%) although the nuclear continuum emission arise from 
more compact region than the NLR emission.
This problem can be completely solved by remembering the
dilution effect by the the starlight of the host galaxy on the
polarization degree of the continuum emission.
We can correct the dilution effect by adopting the fraction of the
starlight in the continuum emission.
If we adopt 84\% as a starlight fraction of the continuum emission
(see \S3), the intrinsic polarization degree of the continuum emission
is calculated to be $1.0 / (1-0.84) \sim 6.3$\%, which is roughly 
consistent with the previous result of Kay (1994).
This corrected polarization degree of the nuclear continuum
emission is significantly larger than the polarization degrees of
all forbidden emission lines.
Note that the difference in the polarization degree between the
NLR emission and the starlight-corrected continuum emission is due 
to the dilution effect by the unobscured part of the NLR.
As mentioned above, some forbidden lines of Mrk 573 have been 
observed to be spatially extended (see, e.g., Unger et al. 1987;
Haniff et al. 1988; Tsvetanov \& Walsh 1992; Pogge \& De Robertis 1995; 
Capetti et al. 1996; Schmitt et al. 2003). Emission-line imaging 
observations with a high quality and a high spatial resolution for 
higher ionization forbidden lines is important to interpret the 
polarization properties of the NLR emission further.


\acknowledgments

We are grateful to all the staffs of the 
Subaru telescope, especially to the FOCAS instrument team.
We thank the referee, R. Goodrich, for his comments which improve
this paper fruitfully.
We also thank K. Matsuda and M. Seki for useful comments.
TN acknowledges financial support from the Japan Society for the
Promotion of Science (JSPS) through JSPS Research Fellowships for Young
Scientists. A part of this work was financially supported by Grants-in-Aid
for the Scientific Research (10044052, 10304013, and 13740122) of the 
Japanese Ministry of Education, Culture, Sports, Science, and Technology.


\end{document}